\documentstyle[12pt]{article}
\def\be {\begin{equation}}
\def\ee {\end{equation}}
\def\ba {\begin{eqnarray}}
\def\ea {\end{eqnarray}}

\def\bi {\begin{itemize}}
\def\ei {\end{itemize}}
\begin{document}
\def\bea{\begin{eqnarray}}
\def\eea{\end{eqnarray}}
\title{\bf { The Generalized Uncertainty
Principle and Corrections to the Cardy-Verlinde Formula in
$SAdS_5$ Black Holes}}
 \author{M.R. Setare  \footnote{E-mail: rezakord@ipm.ir}
  \\{Physics Dept. Inst. for Studies in Theo. Physics and
Mathematics(IPM)}\\
{P. O. Box 19395-5531, Tehran, IRAN }}

\maketitle
\begin{abstract}
In this letter, we  investigate a possible modification to the
temperature and entropy of $5-$dimensional Schwarzschild anti de
Sitter black holes due to incorporating stringy corrections to the
modified uncertainty principle. Then we subsequently argue for
corrections to the Cardy-Verlinde formula in order to account for
the corrected entropy. Then we show, one can taking into account
the generalized uncertainty principle corrections of the
Cardy-Verlinde entropy formula by just redefining the Virasoro
operator $L_0$ and the central charge $c$.
 \end{abstract}
\newpage

 \section{Introduction}
 It is commonly believed that any valid theory of quantum gravity
 must necessary incorporate the Bekenestein-Hawking definition of
 black hole entropy \cite{bek,haw} into its conceptual framework \cite{alen} .
 However, the microscopic origin of this entropy remains an enigma
 for two reasons. First of all although the various counting
 methods have pointed to the expected semi-classical result, there
 is still a lack of recognition as to what degrees of freedom are
 truly being counted. This ambiguity can be attributed to most of
 these methods being based on dualities with simpler theories,
 thus obscuring the physical interpretation from the perspective of
 the black hole in question. Secondly, the vast and varied number
 of successful counting techniques only serve to cloud up an
 already fuzzy picture.\\
  The Cardy-Verlinde formula proposed
   by Verlinde \cite{Verl}, relates the entropy of a  certain CFT with its total
energy and its Casimir energy in arbitrary dimensions. Using the
AdS$_{d}$/CFT$_{d-1}$ \cite{AdS} and dS$_{d}$/CFT$_{d-1}$
correspondences \cite{AS} , this formula has been shown to hold
exactly for different black
holes (see for example {\cite{odi}-\cite{set2}}).\\
Black hole thermodynamic quantities depend on the Hawking
temperature via the usual thermodynamic relations. The Hawking
temperature undergoes corrections from many sources:the quantum
corrections, the self-gravitational corrections, and the
corrections due to the generalized uncertainty principle.\\
It has been known for some time that the quantum effect (a quantum
correction to the microcanonical entropy due to the correction to
the number of microstates, and another correction due to the
thermal fluctuation around equilibrium state) result in
logarithmic corrections to the black hole entropy
\cite{maj1x}-\cite{carli}. \\
Concerning the Hawking effect \cite{hawking1} much work has been
done using a fixed background during the emission process. The
idea of Keski-Vakkuri, Kraus and Wilczek (KKW)
\cite{KKW1}-\hspace{-0.1ex}\cite{KKW3} is to view the black hole
background as dynamical by treating the Hawking radiation as a
tunnelling process. The energy conservation is the key to this
description. The total (ADM) mass is kept fixed while the mass of
the black hole under consideration decreases due to the emitted
radiation. The effect of this modification gives rise to
additional terms in the formulae concerning the known results for
black holes \cite{correction1}-\hspace{-0.1ex}\cite{correction4};
a nonthermal partner to the thermal spectrum of the Hawking
radiation shows up.\\
The generalized uncertainty principle corrections are not tied
down to any specific model of quantum gravity; these corrections
can be derived using arguments from string theory \cite{amati} as
well as other approaches to quantum gravity \cite{magi}.\\
Previous studies of the corrected Cardy-Verlinde formula in
AdS/CFT or dS/CFT context have neglected the corrections due to
the generalized uncertainty principle \cite{set9} In the present
paper, we take into account  corrections to the Cardy-Verlinde
entropy formula of the five-dimensional SAdS black hole that arise
due to the generalized uncertainty principle. In section $2$ we
review the connection between uncertainty principle and
thermodynamic quantities, then we drive the corrections to these
quantities due to the generalized uncertainty principle
\cite{gupdas}. In section $3$ we consider the Cardy-Verlinde
formula of a $5-$dimensional Schwarzschild anti de Sitter black
hole, then we obtain the generalized uncertainty principle
corrections to this entropy formula.

\section{The generalized uncertainty principle}
The metric of an SAdS  black hole in $5-$dimension is given by \be
ds^{2}=-(1-\frac{16\pi
G_{5}M}{3\Omega_{3}c^{2}r^{2}}+\frac{r^{2}}{l^{2}})dt^{2}+
(1-\frac{16\pi
G_{5}M}{3\Omega_{3}c^{2}r^{2}}+\frac{r^{2}}{l^{2}})^{-1}dr^{2}+r^{2}d\Omega_{3}^{2},
\label{metr} \ee where $\Omega_{3}$ is the metric of the unit
$S^{3}$ and $G_{5}$ is the $5-$dimensional Newton's constant.
Since the Hawking radiation is a quantum process, the emitted
quanta must satisfy the Heisenberg uncertainty principle \be
\Delta x_{i}\Delta p_{j}\geq \hbar \delta_{ij}, \label{prin} \ee
where $x_{i}$ and $p_{j}$, $i,j=1...4$, are the spatial
coordinates and momenta, respectively. By modelling a black hole
as a $5-$dimensional cube of size equal to twice its Schwarzschild
radius $r_{+}$, the uncertainty in the position of a Hawking
particle at the emission is \be \Delta x\approx
2r_{+}=l\sqrt{\frac{-1+\sqrt{1+\frac{64\pi
G_{5}M}{3l^{2}\Omega_{3}c^{2}}}}{2}}, \label{delta} \ee
 Using Eq.(\ref{prin}), the uncertainty in the energy
of the emitted particle is \be \Delta E\approx c\Delta
p\approx\frac{\hbar}{l\sqrt{\frac{-1+\sqrt{1+\frac{64\pi
G_{5}M}{3l^{2}\Omega_{3}c^{2}}}}{2}}},\label{dele}\ee The entropy,
 Hawking temperature and energy of black hole are as
 \be S_{BH}=\frac{\Omega_{3}r_{+}^{3}}{4l_{p}^{3}}
 \approx\frac{\Omega_{3}}{4l_{p}^{3}}( \frac{\pi l^{2}}{\hbar c})^{3}T^{3} \label{entro}
\ee
 \be
T=\frac{\hbar c(4r_{+}^{2}+2l^{2})}{4\pi l^{2}r_{+}}\approx
\frac{\hbar c}{\pi l^{2}}r_{+}, \hspace{1cm}r_{+}\gg l,
\label{temp1} \ee \be E=\frac{3\Omega_{3}r_{+}^{2}c^{4}}{16\pi
G_{5}}(1+\frac{r_{+}^{2}}{l^{2}}) \label{ener} \ee where the
approximation $r_{+}\gg l$ is known as the high-temperature limit.
We now determine the corrections to the above results due to the
generalized uncertainty principle. The general form of the
generalized uncertainty principle is \be \Delta x_{i}\geq
\frac{\hbar}{\Delta p_{i}}+\alpha^{2}l_{lp}^{2} \frac{\Delta p_{i}
}{\hbar}, \label{genral}\ee where $l_{pl}=(\frac{\hbar
G_5}{c^{3}})^{1/3}$ is the Planck length and $\alpha$ is a
dimensionless constant of order one. There are many derivations of
the generalized uncertainty principle, some heuristic and some
more rigorous. Eq.(\ref{genral}) can be derived in the context of
string theory \cite{amati}, non-commutative quantum mechanics
\cite{magi}, and from minimum length consideration \cite{gary}.
The exact value of $\alpha$ depends on the specific model. The
second term in r.h.s of Eq.(\ref{genral}) becomes effective when
momentum and length are of the order of Planck mass and of the
Planck length, respectively. This limit is usually called quantum
regime. Inverting Eq.(\ref{genral}), we obtain \be \frac{\Delta
x_{i}
}{2\alpha^{2}l_{pl}^{2}}[1-\sqrt{1-\frac{4\alpha^{2}l_{pl}^{2}}{\Delta
x_{i}^{2} }}]\leq \frac{\Delta p_{i}}{\hbar}\leq \frac{\Delta
x_{i}
}{2\alpha^{2}l_{pl}^{2}}[1+\sqrt{1-\frac{4\alpha^{2}l_{pl}^{2}}{\Delta
x_{i}^{2} }}] \label{momengen} \ee Now we consider the corrections
to the black hole thermodynamic quantities. Setting $\Delta
x=2r_{+}$ and using Eq.(\ref{temp1}) the generalized uncertainty
principle-corrected Hawking temperature is \be T'=\frac{c\hbar
\alpha^{2}l_{pl}^{2}}{2l^{2}\pi
r_{+}(1-\sqrt{1-\frac{\alpha^{2}l_{pl}^{2}}{r_{+}^{2}}})}
\label{tempcorr} \ee Denominator  Eq.(\ref{tempcorr}) may be
Taylor expanded around $\alpha=0$: \be T'= \frac{c\hbar r_{+}
}{\pi
l^{2}(1+\frac{\alpha^{2}l_{pl}^{2}}{4r_{+}^{2}})}=\frac{c\hbar
r_{+} }{\pi
l^{2}}(1-\frac{\alpha^{2}l_{pl}^{2}}{4r_{+}^{2}})=(1-\frac{\alpha^{2}l_{pl}^{2}}{4r_{+}^{2}})T.\label{expan}
\ee The generalized uncertainty principle-corrected Hawking
temperature is smaller than the semiclassical Hawking temperature
$T$ of Eq.(\ref{temp1}).
 The generalized uncertainty principle-corrected black hole
entropy is \be S'_{BH}=\frac{\Omega_{3}}{4l_{p}^{3}}(\frac{\pi
l^{2} }{\hbar
c})^{3}T'^{3}=\frac{\Omega_{3}}{4l_{p}^{3}}(\frac{\pi l^{2}
}{\hbar
c})^{3}(1-\frac{3\alpha^{2}l_{p}^{2}}{4r_{+}^{2}})T^{3}=S_{BH}(1-\frac{3\alpha^{2}l_{p}^{2}}{4r_{+}^{2}}).
\label{corent} \ee From Eq.(\ref{corent}) it follows that the
corrected entropy is smaller than the semiclassical
Bekenstein-Hawking.
\section{Generalized Uncertainty Principle Corrections to the Cardy-Verlinde Formula}
The entropy of a $(1+1)-$dimensional CFT is given by the
well-known Cardy formula \cite{Cardy} \be
S=2\pi\sqrt{\frac{c}{6}(L_0-\frac{c}{24})}, \label{car} \ee where
$L_0$ represent the product $ER$ of the energy and radius, and
the shift of $\frac{c}{24}$ is caused by the Casimir effect. After
making the appropriate identifications for $L_0$ and $c$, the
same Cardy formula is also valid for CFT in arbitrary spacetime
dimensions $d-1$ in the form \cite{Verl} \be S_{CFT}=\frac{2\pi
R}{d-2}\sqrt{E_c(2E-E_c)}, \label{cardy}
 \ee the so called Cardy-Verlinde formula, where $R$ is the radius of the system,
 $E$ is the total energy and $E_c$ is the Casimir
 energy, defined as
 \be E_c=(d-1)E-(d-2)TS.  \label{casi} \ee
 \\
 In this
section we compute the generalized uncertainty principle
corrections to the entropy of a $(d=5)-$dimensional Schwarzschild
anti de Sitter black hole described by the Cardy-Verlinde formula
Eq.(\ref{cardy}). The Casimir energy Eq.(\ref{casi}) now will be
 modified due to the the uncertainty principle corrections as
\be E'_{c}=4E'-3T'S'.  \label{casi1} \ee It is easily seen that
\bea E'_{c}(2E'-E'_{c})&=&(4\pi T'-3T'S'_{BH})(-2\pi
T'+3T'S'_{BH})\nonumber \\
&=& -8\pi^{2}T^{2}-16\pi^{2}T \Delta T+18\pi S_{BH}T^{2}+38\pi
TS_{BH}\Delta T+18\pi T^{2}\Delta S- \nonumber  \\
&&9T^{2}S_{BH}^{2}-18T^{2}S_{BH}\Delta S-18TS_{BH}^{2}\Delta
T.\label{delcasmi}\eea We substitute the previous expression (17)
in the Cardy-Verlinde formula in order that generalized
uncertainty principle corrections to be considered, \bea
\label{cvcor} S'_{CFT}=S_{CFT}[1+\frac{T[-16\pi^{2}\Delta T+18 \pi
T \Delta S+30\pi S_{BH}\Delta T-18 S_{BH} T \Delta S-18
S_{BH}^{2}\Delta T]}{2E_c(2E-E_c)}] \eea where \be \Delta
T=\frac{-\alpha^{2}l_{p}^{2}}{4r_{+}^{2}}T \label{tempcor} \ee \be
\Delta S=\frac{-3\alpha^{2}l_{p}^{2}}{4r_{+}^{2}}S_{BH}
\label{delent}\ee If we would like to express the modified
Cardy-Verlinde entropy formula in terms of the energy and Casimir
energy, it is necessary to rewrite the $T, S_{BH}, \Delta T,
\Delta S $ in terms of energy as following \be T=\frac{2
\hbar}{\pi l^{2}}(\frac{\pi G_{5}l^{2}E}{3\Omega_{3}})^{1/4}
\label {tem1} \ee \be S_{BH}=\frac{\Omega_{3}c^{3}}{4\hbar
G_5}(\frac{16 \pi G_5 l^{2}E}{3\Omega_{3}c^{4}})^{3/4}
,\label{ent1} \ee \be \Delta T=\frac{-\alpha^{2}\hbar
c^{2}l_{p}^{2}}{8 \pi l^{2}}(\frac{3\Omega_{3}}{\pi G_5
l^{2}E})^{1/4} , \label{temcor1} \ee \be \Delta
S=\frac{-3\alpha^{2}\Omega_{3}}{16 l_{p}}(\frac{16\pi G_5 l^{2}
E}{3\Omega_{3}c^{4}})^{1/4} \ee As we saw in above discussion
these corrections are
caused by generalized uncertainty principle.\\
Then, we can taking into account the generalized uncertainty
principle corrections of the Cardy-Verlinde entropy formula by
just redefining the Virasoro operator, $L_0=ER$, and the central
charge $\frac{c}{6}=\frac{(d-2)S_c}{\pi}=2E_cR$, where $S_c$ is
the Casimir entropy \be
L'_{0}=E'R=\frac{3\Omega_{3}l^{6}\pi^{3}}{16\hbar^{4}G_5}T'^{4}R=(1-\frac{\alpha^{2}l_{p}^{2}}{r_{+}^{2}})ER=
(1-\frac{\alpha^{2}l_{p}^{2}}{r_{+}^{2}})L_0 \label{LEQ} \ee \bea
c'=12E'_{c}R=12(4E'-3S'_{BH}T')R=12R(1-\frac{\alpha^{2}l_{p}^{2}}{r_{+}^{2}})E_c=(1-\frac{\alpha^{2}l_{p}^{2}}
{r_{+}^{2}})c \label{LEQ1} \eea his redefinition includes only a
multiplicative constant
 term and, therefore can be considered as a renormalization of the quantities entering in the Cardy formula.
 \footnote{To see similar redefinition of the $c$ and $L_0$ in the Cardy formula due to the quantum corrections
 refer to the \cite{serg} }In \cite{carx} Carlip have computed the
 logarithmic corrections to the Cardy formula, according to his
 calculations, logarithmic corrections to the density of states is
 as \be \rho(\Delta)\approx (\frac{c}{96\Delta^{3}})^{1/4}\exp (2\pi\sqrt{\frac{c\Delta}{6}}), \label{carcor} \ee
where $\Delta=L_0$, the exponential term in (\ref{carcor}) gives
the standard Cardy formula, but we have now found the lading
correction, which is logarithmic. In the other hand as we saw the
 effect of the generalized uncertainty principle to the Cardy-
 Verlinde formula appear as the redefinition of the $c$ and $L_0$
 only. As Carlip have discussed in \cite{carx}, the central charge
 $c$ appearing in (\ref{carcor}) is the full central charge of the
 conformal field theory. In general, $c$ will consist of a
 classical term which appear in the Poisson bracets of the
 Virasoro algebra  generators, plus a correction due to the
 quantum (here generalized uncertainty principle) effects, that
 can change the exponent in (\ref{carcor}) from its classical
 value. Moreover we saw that $L_0$ take a similar correction as
 eq.(\ref{LEQ}), then the similar discussion about $L_0$ is
 correct.
\\Therefore the first order corrections to the $L_0$ and $c$ are
given by \be \Delta
L_0=L'_{0}-L_0=(E'-E)R=\frac{-\alpha^{2}l_{p}^{2}}{r_{+}^{2}}L_{0}
 \label{l1eq}
\ee \bea\Delta
c&=&c'-c=12R(E'_{c}-E_{c})=\frac{-\alpha^{2}l_{p}^{2}}{r_{+}^{2}}c
\label{cendel}
 \eea

  \vspace{3mm}

\section{Conclusion}
In this paper we have examined the effects of the generalized
uncertainty principle in the generalized Cardy-Verlinde formula.
The general form of the generalized uncertainty principle is
given by Eq.(\ref{genral}). Black hole thermodynamic quantities
depend on the Hawking temperature via the usual thermodynamic
relation. The Hawking temperature undergoes corrections from the
generalized uncertainty principle as Eq.(\ref{tempcorr}). Then we
have obtained the corrections to the entropy of a dual conformal
field theory live on boundary space as Eq.(\ref{cvcor}). Then we
have considered this point that the Cardy-Verlinde  formula is the
outcome of a striking resemblance between the thermodynamics of
CFTs with asymptotically Ads dual's and CFTs in two dimensions.
After that we have obtained the corrections to the quantities
entering the Cardy-Verlinde formula:Virasoro operator and the
central charge.

\end{document}